\documentclass[12pt]{article}

\usepackage{graphicx}

\begin{document}

\begin{center}
{\Large\bf \boldmath Nuclear fusion in muonic molecules and
in deuterated metals} 

\vspace*{6mm}
{L.N.~Bogdanova}\\      
{\small \it State Scientific Center of RF Institute for
Theoretical and Experimental Physics}
\end{center}

\vspace*{6mm}

\begin{abstract}
Study of the fusion reactions between hydrogen isotopes in muonic
molecules is the first example of the accurate accounting of the
nucleus charge screening by a muon in the fusion process. At LUNA
installation the measurements of astrophysical reaction cross
sections were extended down to collision energies of a few keV.
The screening by atomic electrons of the target became
substantial. The possibility to look over screening from unbound
electrons is given by metal-hydrides used as targets in $dd$
reaction measurements. The classical Debye screening in plasma,
applied to quasi-free electrons in metal, provides an explanation
of unexpectedly large screening potentials found for some metals
in the research through the Periodic table of elements.

\end{abstract}

\vspace*{6mm}

\section {Electron and muon screening of fusion reactions}
 Fusion reaction cross sections decrease exponentially
with decreasing of collision energy. It is convenient to isolate
this dependency and characterize the reaction with the
astrophysical function $S(E)$ rather than with the reaction cross
section $\sigma(E)$:
\begin{equation}
S(E)= E \sigma (E)exp(2 \pi \eta),
\end{equation}
where $\eta$ is the Sommerfeld parameter, $2\pi \eta = 31.29 Z_1
Z_2 (m/E)^{0.5}$, $Z_1$ and $Z_2$ are charges of colliding nuclei,
$m$ is their reduced mass in amu, $E$ is the c.m. collision energy
in keV ($\hbar=c=1$). Usually $S(E)$ is a smooth function of
energy. However, the success in the extrapolation of the measured
$S$-function to the energy of the astrophysical scale is not
guaranteed: first, due to possible occurrence of bound or resonant
states at low energies, second, due to the effect from atomic
electrons of the beam or the target in collision experiments
\cite{Assen}. The latter effect leads to the increase of the cross
section in comparison with the cross section of the interaction
between "bare" nuclei which is expressed in terms of an
enhancement factor
\begin{equation}
f(E) = \sigma_{\rm exp} (E) / \sigma_{\rm bare} (E),
\end{equation}
where $\sigma_{\rm bare}$ and $\sigma_{\rm exp}$ are the cross
sections for bare nuclei and the one measured in the presence of
target and projectile atomic electrons.

Electrons can be thought of as effectively lowering the Coulomb
energy between the two colliding nuclei by a {\it constant and
energy-independent} increment $U_e$, the {\it screening
potential}, which is usually treated as the acceleration of the
incident particle by the electron cloud. In the adiabatic limit,
the accelerating potential, $U_{\rm ad}$, is evaluated from the
difference in atomic binding energies between the compound atom
and the projectile plus the target atoms of the entrance channel.
For reaction $D(D,p)t$  the accelerating potential is equal to
$U_{\rm ad}=2\cdot13.6~{\mbox eV}=27.2$~eV due to atomic electrons
at the Bohr radius. The electron screening effect was observed for
reactions $d(d,p)t$ and $d(d,n)^3{\mbox He}$ with a molecular
target in \cite{Greife}. The extracted magnitude of the screening
potential,
\begin{equation}
 U_e^{gas} = 25 \pm 5 ~{\mbox eV},
\end{equation}
 qualitatively agreed with the above estimate (Fig. 1).

 At lowest energies currently
accessible in the laboratory, 1-10 keV, $E \gg U_e $, the
enhancement factor behaves exponentially
 \begin{equation}
f(E)= \sigma_{\rm bare}(E+U_e)/\sigma_{\rm bare}(E) \approx
exp(\pi \eta U_e /E)\geq 1
\end{equation}
and even a small increase of the cross section at $E/U_e\approx
100$ can lead to a wrong extrapolation if the screening correction
is not properly introduced.
\begin{figure}[h]
  \hfill
  \begin{minipage}[t]{.45\textwidth}
    \begin{center}
      \includegraphics[width=5.5cm, angle=90]{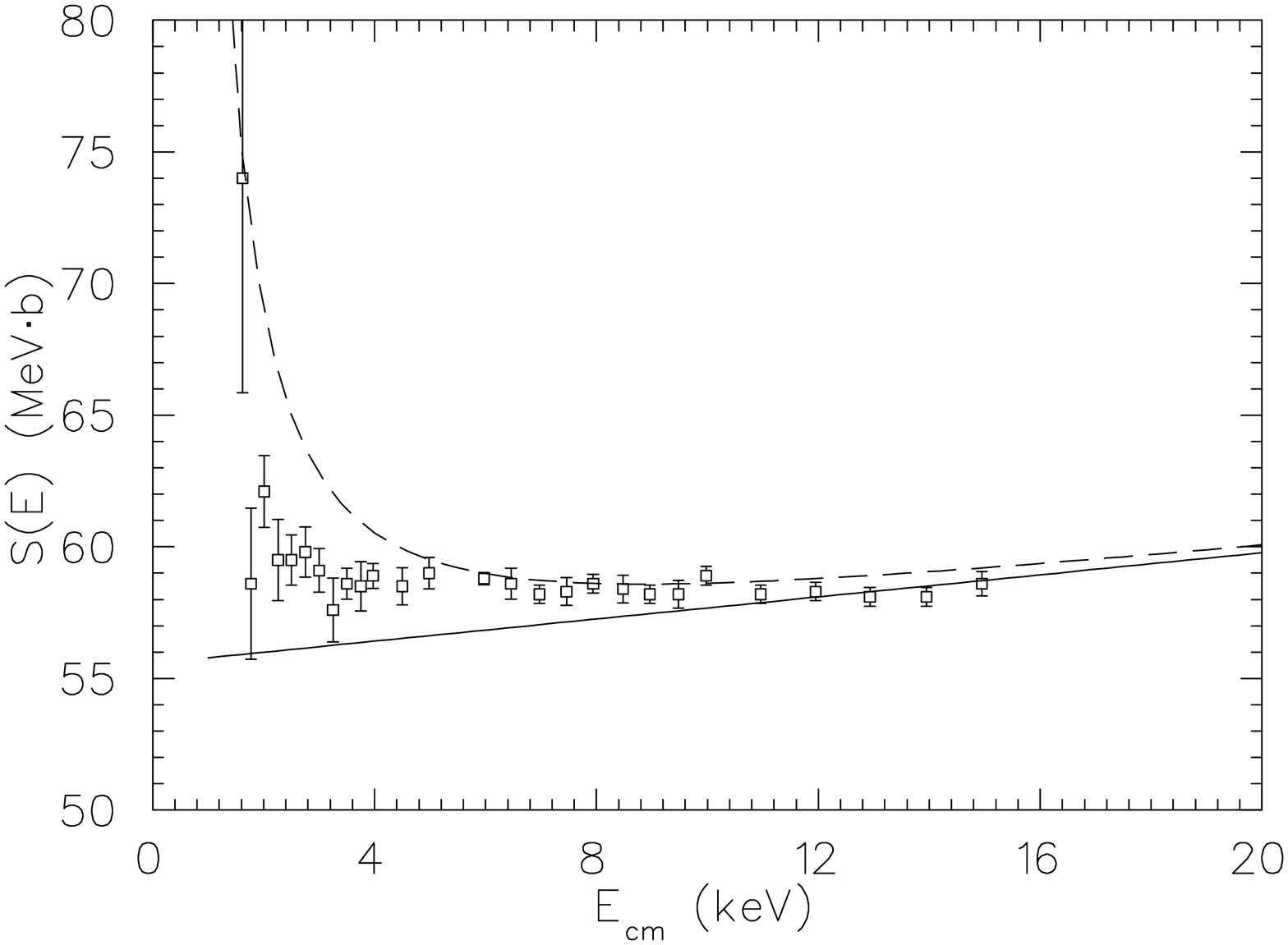}
      \caption{Comparison of the experimental $d(d,p)t~~~~S(E)$-functions
\cite{Greife} with the bare-nuclei $S(E)$-function from Ref.
\cite{Hale} (solid line) and estimate \cite{Shoppa} (dashed line)
in which electron screening is added to the bare-nuclei
$S$-function using Eq. (4).}
    \end{center}
  \end{minipage}
  \hfill
  \begin{minipage}[t]{.45\textwidth}
    \begin{center}
      \includegraphics[width=8cm]{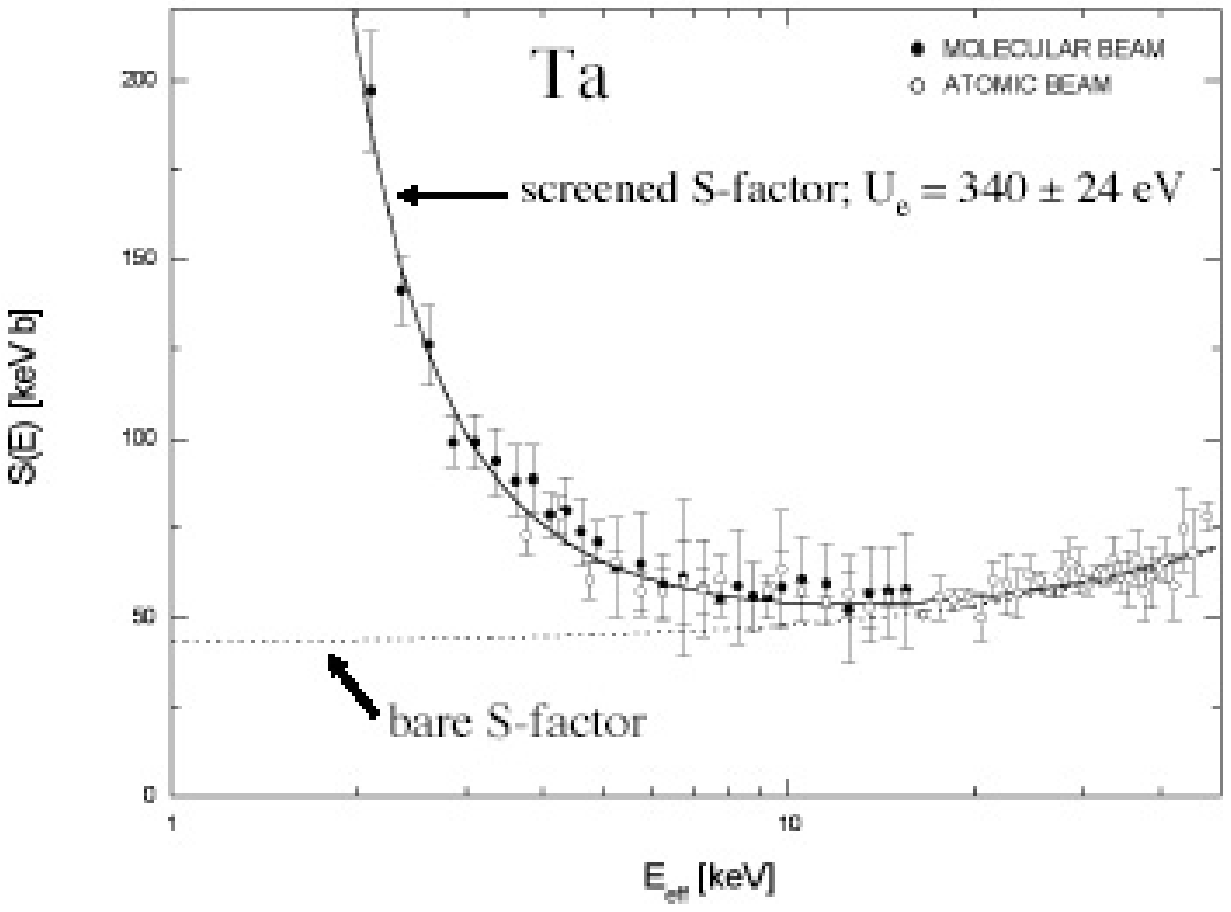}
      \caption{Measured astrophysical $S(E)$-function of $d(d,p)t$ obtained with
a deuterated Ta foil (T=-10$^\circ$C) for atomic and molecular
deuteron beams \cite{Ta}. The dotted curve represents the bare
$S(E)$-function, the solid one includes the effects of electron
screening with $U_e$=340~eV.}
    \end{center}
  \end{minipage}
  \hfill
\end{figure}

In order to reduce the uncertainties behind extrapolation
procedures in collision experiments, considerable efforts have
been spent to push the experimental limits towards lower energies.
LUNA collaboration measured the cross sections of the reactions
$^3{\mbox He}(d,p)^4{\mbox He}$ and $d(^3{\mbox He},p)^4{\mbox
He}$ at the c.m. energies, respectively, $E$ = 5 to 60 keV and 10
to 40 keV and studied the screening effect  \cite{aliotta}. The
magnitude of screening potentials for both reactions, $U_e = 219
\pm 7$ and $109 \pm 9$ eV, appeared to be significantly larger
than the values from atomic physics models, $U_e = 120$ and 65 eV.
This is not understood so far. The experimental analysis generally
assumes that the bare-nuclei cross section, e.g., \cite{Hale},
derived by the extrapolation of data at higher energies, which are
little affected by screening, is known. Authors \cite{aliotta}
conclude that, besides additional theoretical work on screening, a
direct measurement of bare-nuclei $S(E)$-function would be
desirable to clear up the situation.

 A possibility to verify the extrapolation is
given by the muon catalyzed fusion. Muonic molecule is an ideal
model system, which makes possible study of fusion reactions
screened by a muon at practically zero energy \cite {Bogdanova}.
The rates of nuclear fusion from the muonic molecule bound
rotational-vibrational states $(Jv)$, $\lambda^{Jv}_L$,  are
expressed through the "bare" {\em reaction constants} $K_{L}$ or
the "bare" $S$-functions
\begin{equation}
 \lambda^{Jv}_L = K_{L} \cdot G^{Jv}_L, ~~~~~K_L=S_L(0)/(\pi (\alpha
 m)^{2L+1}).
\end{equation}
Here $S_L$ is the $S$-function, corresponding to the $L$-wave
partial cross section,  $L$ is the orbital angular momentum of
nuclei, ~$G^{Jv}_L$ is the density probability of nuclei
penetration to the fusion region. This quantity, the "muon
screening factor", was accurately calculated for muonuc molecules
(see \cite {local} and refs. therein).

At PSI, radiative capture rates from the ground $(00)$ state of
$pd\mu$ muonic molecule were measured for doublet and quartet
nuclear spin states \cite {Peti}. At TUNL, doublet and quartet
cross sections of reactions $p(\vec{d},\gamma)^3\!{\mbox He}$ and
$d(\vec{p},\gamma)^3\!{\mbox He}$ with polarized nuclei were
obtained for energies down to $E_{cm}\cong $ 23 keV and
extrapolated to $E=0$ \cite{Rice}. The values $S_0(0)$ for the
s-wave deduced from these experiments: $S_0(0)=0.109 \pm 0.01$
\cite {Rice} and $S_0(0)=0.105 \pm 0.01$ \cite{Peti} (in units of
eV b), are in an agreement, the extrapolation being thus verified.

 LUNA collaboration measured the $d(p,\gamma)^3{\mbox He}$
cross section from 22 down to 2.5 keV c.m. energy, well below the
solar Gamow peak \cite {LUNA_pd}, where the cross section falls to
values about picobarns. Their result for the total (($s+p$)-wave)
extrapolated $S$-function $S(0)= S_0 +S_1 = 0.216 \pm 0.010 $ is
somewhat larger than that deduced from \cite {Rice} $S(0)= 0.166
\pm 0.014 $. The difference is within the uncertainty due to
electron screening effect: the estimated enhancement of the
S-factor is about 6\% at 2.5 keV (c.m.) and it increases to ~20\%
for interaction energies around 1 keV, but screening potential
could not be extracted from the data \cite {LUNA_pd}.

Characteristic features of charge-symmetric reactions
$d(d,n)^3{\mbox He}$ and $d(d,p)^3{\mbox H}$: ratio of channels in
$p$- and $s$-waves and $p$-wave contribution to the total cross
section in both channels were studied via muon catalysis. Ratio
$R(n/p)$ of neutron and proton fusion yields in the $dd\mu$
molecule was measured as a function of deuterium mixture
temperature \cite{Peti}. In conditions of resonant $dd\mu$
formation ($p$-wave fusion) the value $R_p(n/p)=1.42 \pm 0.03$ was
obtained. This ratio $R(n/p)$ decreases to $R_s=1.06 \pm 0.04$,
when non-resonant $dd\mu$ formation and $s$-wave fusion become
prevailing. The $p$-wave reaction constants for mirror channels
were extracted from the fusion rates determined in \cite{Peti}.
All results agree with the R-matrix analysis of $dd$-fusion
in-flight data \cite{Hale}, proving the efficiency of $\mu$CF
approach.

Properties of $dd$ reaction were studied in \cite{Czerski} at a
beam energy as low as 5 keV and target deuterons embedded in the
metallic environment. The angular distributions of both $^2{\mbox
H}(d,p)t$ and $^2{\mbox H}(d,n)^3{\mbox He}$ reactions were
measured. They are expected to be anisotropic even at the lowest
energies due to a relatively large p-wave component in the
entrance channel. The angular distribution is symmetric with
respect to 90$^o$ and can be parametrized as $A(\theta) = 1 + a_2
cos^2(\theta)$ (neglecting an L=2 contribution). The energy
dependence of the parameter $a_2$ for both reactions recommended
in \cite {Brown} was confirmed at extremely low energies.  This
agrees with the parametrization of \cite {Brown}, with the
experimental results \cite {Cecil} and \cite {Greife}, and with
the $\mu CF$ result \cite{Peti}.

The $\mu$CF method might be helpful for understanding electron
screening in $^3{\mbox He}(d,p)^4{\mbox He}$ and $d(^3{\mbox
He},p)^4{\mbox He}$ nuclear reactions discussed above. The fusion
rate in muonic molecule $d\mu^3{\mbox He}$ was calculated using
advanced methods of solving three-body Colomb problem \cite{BGP}.
Its measurement would provide the "bare" reaction constant, which
is necessary for adequate fitting of the in-flight data
\cite{aliotta}.

\section{Electron screening effects in solids}
 The application of metal hydrides  as targets provides a unique
possibility to investigate the influence of free (unbound)
electrons on fusion cross sections in beam-target collisions.
During past decade several groups have performed a series of
experiments bombarding deuterated metals with low energy
deuterons. The electron screening effect in $d(d,p)t$ reaction for
the target produced via implantation of low-energy deuterons
unexpectedly appeared to be about one order of magnitude larger
than the value (3) from a gas target experiment \cite{Greife}. As
an example, Fig. 2 from \cite{Ta} shows the astrophysical
$S$-function of the $d(d,p)t$ reaction measured in deuterated Ta.
The measured points clearly show the exponential increase at low
energies. The derived screening potential for Ta is $U_e = 340\pm
24$~eV, which is significantly larger than the theoretical
predictions based on the adiabatic approach.

Early studies of deutron-induced in-flight reactions on deuterated
targets were initiated because of the interest to cold fusion
experiments. An anomalous enhancement of the fusion yield was
reported \cite{Yuki} for deuterated Pd ($U_e = 250\pm15$ eV) and a
deuterated Au/Pd/PdO multilayer ($U_e = 601 \pm 23$ eV), while
deuterated Ti and Au exhibited a normal ("gaseous") enhancement:
$U_e = 36 \pm 11$ and 23$\pm11$ eV, respectively. \footnote {The
screening effects found for deuterated Pd-containing targets could
not justify cold fusion results.}

 In \cite{Czerski} thick target yields of the fusion reactions
$^2{\mbox H}(d,p)^3{\mbox H}$ and $^2{\mbox H}(d,n)^3{\mbox He}$
were measured on deuterons implanted in three different metal
targets (Al, Zr and Ta) for beam energies ranging from 5 to 60~keV
(Fig. 3). The values of the screening energy also demonstrated a
clear target material dependence
 $U_{sc} = (190\pm15)$~eV
for Al, (297$\pm$8)~eV for Zr and (322$\pm$15)~eV for Ta.
\footnote {At the same time, no target material dependence of the
angular distribution was observed in \cite{Czerski}. Thus the
screening effect noticeable for beam energies below 25 keV should
be equal for both entrance wave function contributions $L = 0$ and
$L = 1$ and should not change the relative intensities of both
contributions compared to the gas target experiments. The ratio of
the total yields of mirror reactions appeared close to unit for
all three metal targets at all energies down to 5 keV, in the
agreement with low-energy extrapolations and $\mu$CF result. Cold
fusion observations are evidently contradicting to this firmly
established feature.}

 Study by LUNA collaboration \cite{Ta}
of the reaction $d(d,p)t$ for deuterated Ta tested surprising
results on large screening energy of \cite{Czerski} with somewhat
different experimental approaches and  confirmed them. This
stimulated systematic research through the Periodic table and
results for several metals, insulators, and semiconductors were
reported in \cite{Ra_02PL} and completed in \cite {Bonomo},
\cite{Raiola_04} (Fig. 4).

\begin{figure}[h]
  \hfill
  \begin{minipage}[t]{.45\textwidth}
    \begin{center}
      \includegraphics[width=7cm]{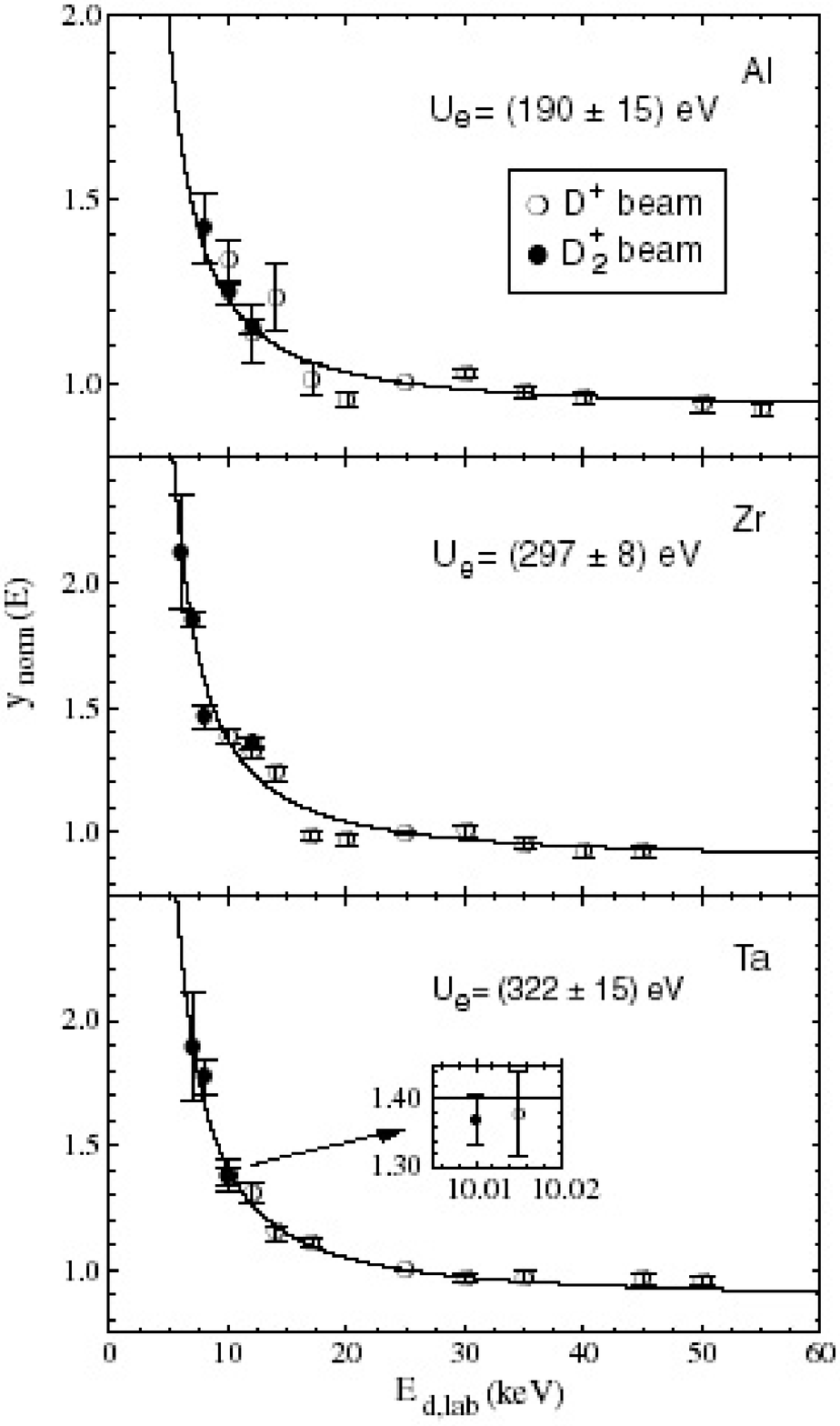} \caption{
Enhancement of the thick-target yields for three different
deuterated metals Al, Zr and Ta from \cite{Czerski} ($E=$ deuteron
lab. energy). Extracted values of the screening potential are
shown.}
    \end{center}
  \end{minipage}
  \hfill
  \begin{minipage}[t]{.45\textwidth}
    \begin{center}
      \includegraphics[width=7cm]{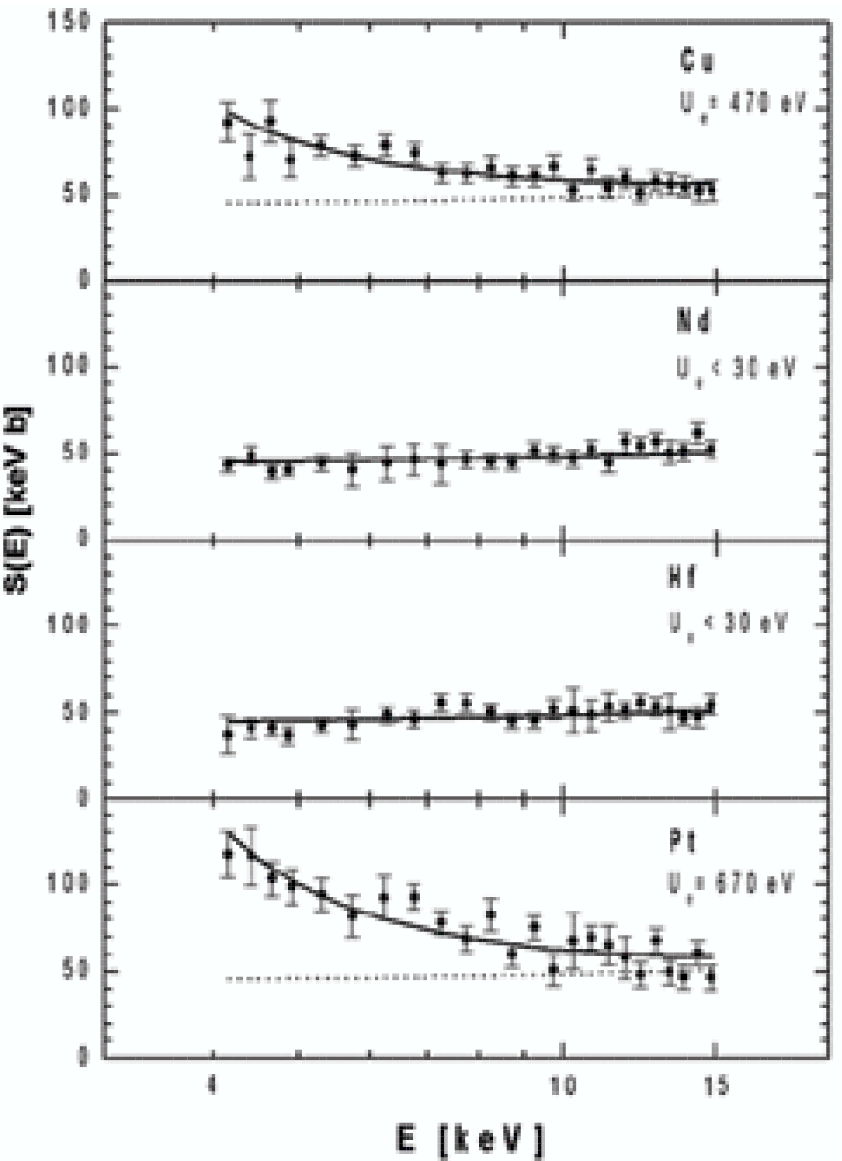} \caption{
      $S(E)$-function of  $d(d,p)t$ reaction
      on deuterated Cu, Nd, Hf, Pt ($E$ in c.m.s.) \cite{Raiola_04}.
      The dotted curve shows the bare
$S(E)$-function, the solid one reproduces the exponential
enhancement with electron screening potential $U_e$.}
    \end{center}
  \end{minipage}
  \hfill
\end{figure}

\begin{figure}[h]
\centerline{\includegraphics[width=13cm]{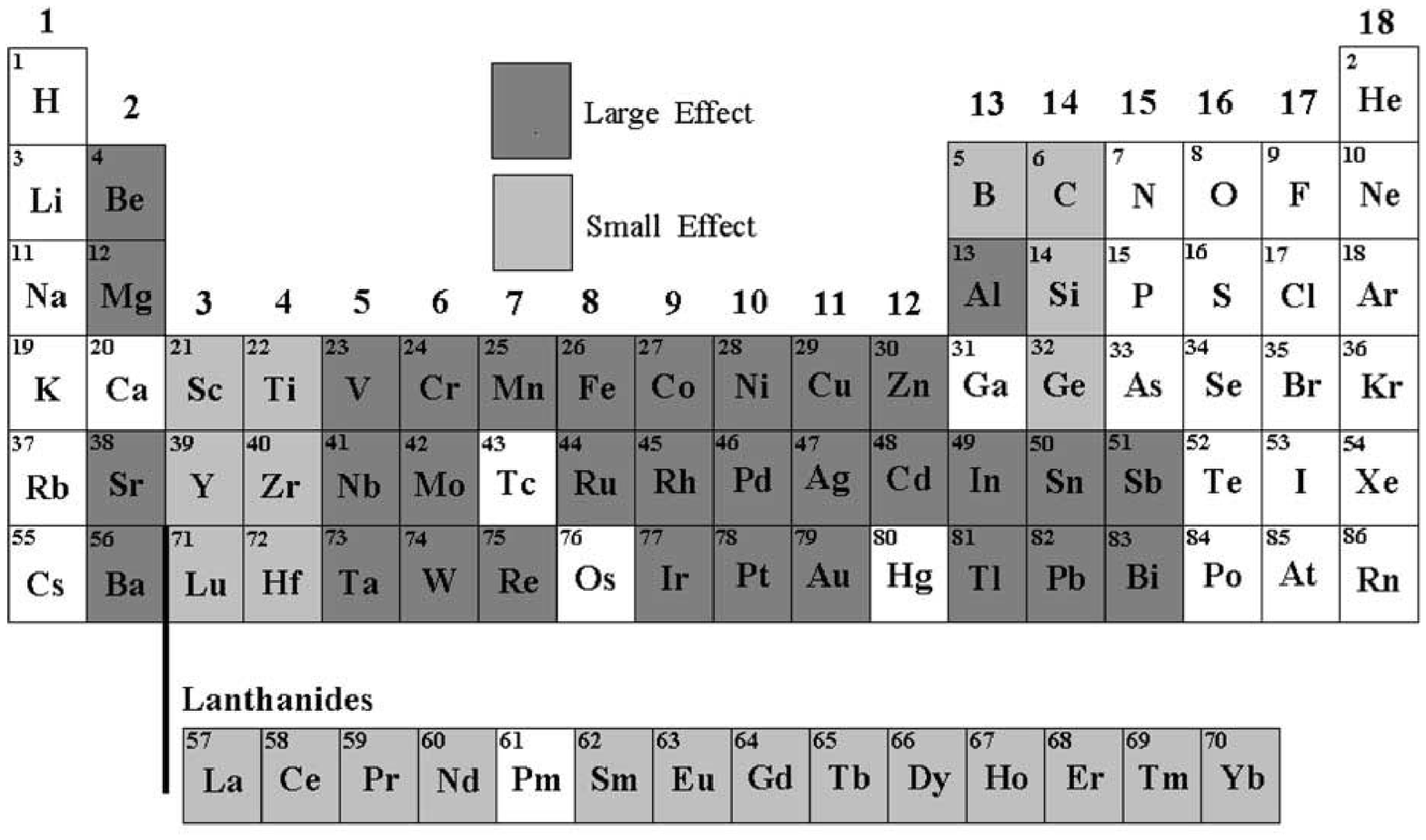}} \caption{
Periodic table showing the studied elements, where those with low
$U_e$ values ($U_e < 100$ eV, small effect) are lightly shadowed
and those with high $U_e$ values ($U_e \geq 100$ eV, large effect)
are heavily shadowed \cite{Raiola_04}.}
\end{figure}

A comparison of the $U_e$ values with the Periodic table indicates
a common attribute: for each group of the Periodic table, the
corresponding $U_e$ values are either low ("gaseous") as for
groups 3 and 4 and the lanthanides, or high as for groups 2, 5 to
12, and 15 (Fig. 5). Group 14 is an apparent exception to this
characteristic: the metals Sn and Pb have a high $U_e$ value,
while the semiconductors C, Si, and Ge have a low $U_e$ value
indicating that high $U_e$ values are a feature of metals. A
similar situation is found for group 13: B = insulator, Al and Tl
= metals. The indication is supported further by the insulators
BeO, Al$_2$O$_3$, and CaO$_2$, as well as by deuterated metals M
having an observed small stoichiometric $x$ value ($M_xD$) of the
order of one or smaller and thus representing also insulators
(e.g. group 4 of the Periodic table and the lanthanides). In
summary, a large screening effect is observed in all metals except
in the noble metals Cu, Ag, and Au.

Various aspects of the metals were discussed to explain the data.
In particular, effects from deuteron thermal and vibrational
motion in the lattice was found to be marginal \cite {Fiorentini}.
Other mechanisms considered in \cite{Ta}, \cite{Ra_02PL} -
channeling, diffusion, conductivity, crystal structure, electron
configuration, and "Fermi shuttle" acceleration - were
unsuccessful either.

The observation that large enhancements have been found in
deuterated metals but not in insulators has suggested a possible
explanation based on effects of the plasma of electrons in the
metal \cite{Bonomo}, \cite{Raiola_04}. This model with classical
quasi-free valence electrons predicts an electron screening
distance around positive singly charged ions (deuterons in the
lattice) of the order of the Debye length
 \begin{equation}
R_D = (kT/(4\pi e^2 n_{eff}\rho_a))^{1/2},
\end{equation}
 where $T$ is the temperature of free electrons,
 $n_{eff}$ is the effective number of valence electrons per metallic atom,
$\rho_a$  is the atomic density.
  At room temperature and at $n_{eff}\approx 1$
 one obtains a radius $R_D$, which is about a factor of 10 smaller
 than the Bohr radius of a hydrogen atom. With the Coulomb energy
 between two deuterons at $R_D$ set equal to $U_e$, one obtains $U_e$ = 300 eV,
 reproducing the correct size of the screening potential $U_e$.
The acceleration mechanism of the incident ions leading to the
high observed $U_e$ values is, within this simple model, the Debye
electron cloud at rather small radius $R_D$.  The $n_{eff}$ values
were compared with those deduced from the known Hall coefficient:
$C_{Hall} = (en_{eff}(Hall)\rho_a)^{-1}$ \cite{Hall}: within 2
standard deviations the two quantities agreed for all metals.

Some consequences of the Debye screening model were tested
experimentally: \begin{itemize}
\item { Debye screening energy $U_D$ is proportional to target nucleus charge.
Tested for pure metals $Z=3$ (Li), 4 (Be), 23 (V), 71 (Lu) \cite
{Li6-7}, \cite {Kettner}.}

\item {The Debye screening energy $U_D$ is proportional to the
nuclear charge of the incident ion. This prediction was verified
\cite{Raiola_04} in the $d(^3He,p)^4He$ studies in metals ($Z_i$ =
2): with a typical value of $U_e$ = 300 eV for the $dd$ fusion in
metals at T = 290 K one would expect for $d(^3He,p)^4He$ the Debye
value $U_D = Z_iU_e(dd) = 2 \times 300$ eV = 600~eV, $U_e = U_{\rm
ad} + U_D = 120 + 600$ eV = 720~ eV, $U_{\rm ad}$ being the
adiabatic screening potential; observation \cite {Raiola_04} gives
$U_e = 680 \pm 60$~ eV.}

\item {The electron screening is the effect of the entrance
channel. The two mirror reactions measured with neutrons and
protons in the exit channel \cite {Czerski}, \cite{Kettner}
demonstrate that the electron screening is not influenced by the
the charged ejectiles of the exit channel.}
\end{itemize}

A critical test of the classical Debye model would be the
predicted screening potential dependence on the temperature: $U_e
\propto T^{-1/2}$. The electron screening in the $d(d,p)t$
reaction has been studied for the deuterated metal Pt at a sample
temperature T = 20$^o$C $\div 340^o$C and for Co at T = 20$^o$C
and 200$^o$C \cite {Raiola_05_G}. The astrophysical
$S(E)$-function obtained at T = 20$^o$C and 300$^o$C for Pt is
shown in Fig. 6. The $U_e(T)$ values for Pt are plotted in Fig. 7
together with the expected dependence $U_e \propto T^{-1/2}$
(dotted curve). All data show a decrease of the screening, i.e.
the $U_e$ value, with increasing temperature.

\begin{figure}[h]
  \hfill
  \begin{minipage}[t]{.45\textwidth}
    \begin{center}
      \includegraphics[width=7cm]{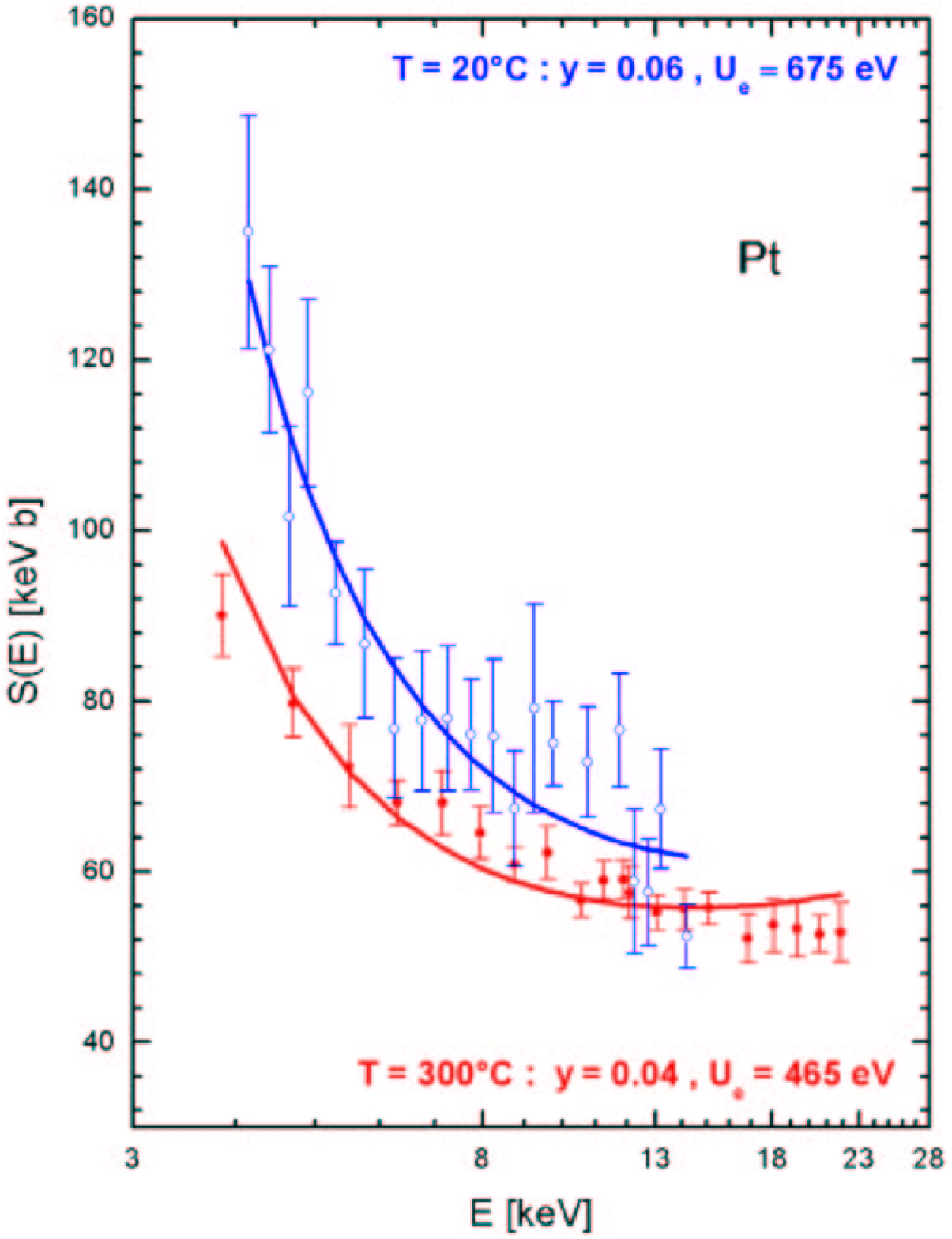}
      \caption{
The $S(E)$-function of $d(d,p)t$ for Pt at T = 20$^o$C and
300$^o$C, with the deduced solubilities $y$. The curves through
the data points include the bare $S(E)$ factor and the electron
screening with the given $U_e$ values \cite {Raiola_05_G}.}
    \end{center}
  \end{minipage}
  \hfill
  \begin{minipage}[t]{.45\textwidth}
    \begin{center}
      \includegraphics[width=7cm]{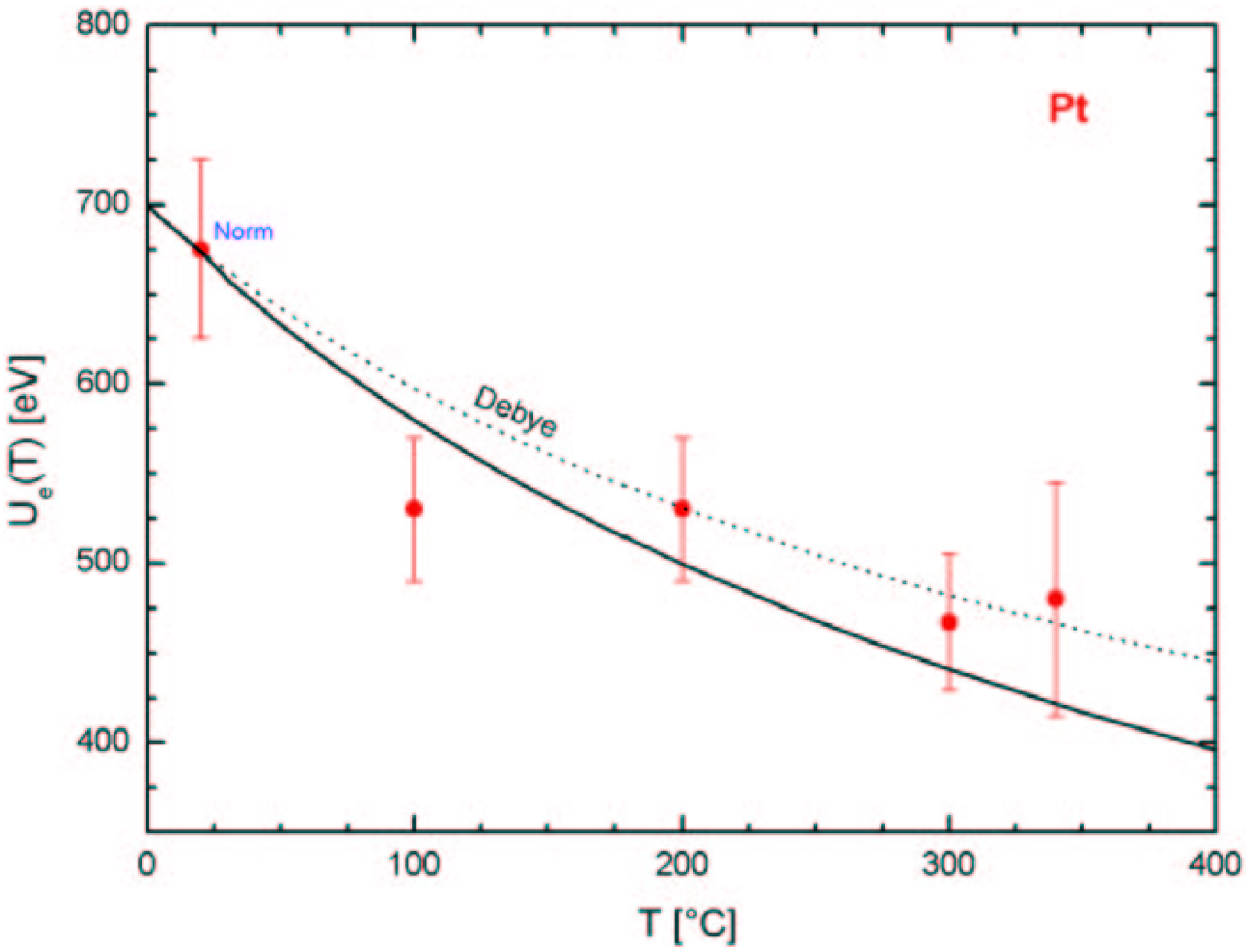} \caption{
 The observed values $U_e(T)$ for Pt are shown as a function of
 sample temperature T. The dotted curve represents the Debye model prediction,
  the solid curve includes the observed
 T-dependence of the Hall coefficient \cite {Raiola_05_G}.}
    \end{center}
  \end{minipage}
  \hfill
\end{figure}

It was noted in \cite {Raiola_04} that the metals with a high
hydrogen solubility, of the order of one, have small screening
potentials. Contrary, for the metals with high $U_e$ values, the
solubilities are quite small. Actual solubilities were not always
available at room temperature, but those observed in
\cite{Raiola_04} were consistent with known results \cite{sprav}.

The temperature dependence of this correlation was further
studied. The electron screening effect for deuterated metals with
high solubility, such as Ti (group 4), was measured at
temperatures from 10 to 200$^o$C. The deduced solubility $y(T)$
shows a sizable decrease with increasing temperature. Above
50$^o$C the solubility reaches values below 1 and an enhanced
screening becomes observable. Finally, all metals of groups 3 and
4 and the lanthanides have been studied at T=200$^o$C.  Similarly
as for Ti, all these metals exhibited a large reduction in
solubility and thus showed a large screening. \footnote {In the
insulator C the solubility decreased to 0.15 (T=200$^o$C) from
0.35 (T = 20$^o$C), but no enhanced screening was observed, as
expected for an insulator with $n_{eff}=0$.} Screening potentials
$U_e$ and  solubilities $y$  for  49 metals in total were measured
in \cite {Raiola_04}, \cite {Raiola_05_G} (see also
\cite{kasagi}).

\section{Discussion}

All data on the enhanced electron screening in deuterated metals
can be explained quantitatively by the Debye model applied to the
quasi-free metallic electrons. So far there is no idea why the
simple Debye model appears to work so well in this case. It was
argued \cite {Raiola_04} that most of the conduction electrons are
frozen by quantum effects and only electrons close to the Fermi
energy ($E_F$) actually should contribute to screening, with
$n_{eff}(T) \sim kT/E_F \propto T $ and thus there should be
essentially no temperature dependence for $U_e$. However, this
argument applies only to insulators and semiconductors with a
finite energy gap, while for metals there is no energy gap and the
Fermi energy lies within the conduction band.

Possible many-body effects characteristic for dense matter
resulting from internuclear correlation processes can distort the
picture; but, due to trapping deuterons by the crystal lattice,
these effects can be neglected. One should refer to the series of
papers by Ichimaru {\em et al.} on nuclear fusion in various types
of dense plasmas (see the review \cite{ichimaru}). In particular,
different effects in metal deuterides were considered. They found
that the electrons both in metallic d-band and hydrogen-induced
s-band contribute to screening and it is more efficient in Pd than
in Ti. However, the magnitude of the screening potential later
observed in \cite {kasagi} remains unexplained, as well as its
temperature dependence as in \cite {Raiola_05_G}.

Without theoretical grounds, the Debye model can be treated as a
powerful parametrization of the data with an excellent predictive
power.

Authors \cite{kasagi} noted  the correlation between the screening
energy and the deuteron density in the host during the
bombardment. They relate the density to the diffusivity, or
mobility of D$^+$ ions in the host; large mobility results in
small density. The inverse of density is related to a concept they
call the "fluidity" of the target deuterons. They propose that
high fluidity in the host might be responsible for the enhanced
screening. The fluid deuterons and conduction electrons might
behave like a plasma in the host and both electrons and positive
ions reduce the Coulomb repulsion. A model of \cite {fukai} is
cited in this context. It is argued that the screening energy due
to the fluid deuterons can be one order of magnitude larger than
that due to the electrons because of the difference between Boson
(deuteron) and Fermion (electron) statistics. Thus, another
dynamic screening mechanism during the deuteron bombardment and
penetration into the host is suggested, wherein the fluidity of
deuterons must play a decisive role.

To explain the observed fusion rate enhancement an approach based
on the quantum-uncertainty dispersive effect between energy and
momentum proposed by Galitskij and Yakimets \cite {G_Y} has also
been applied. Authors \cite {Starostin} showed that the
quantum-tail effect produces an important increment of the rate
enhancement at very low energies, however, this mechanism cannot
reproduce the observed experimental rate for deuterium.

We conclude with the remark that the influence of the environment
in various nuclear processes is one of the interesting subjects
because of its interdisciplinary nature involving nuclear physics,
condensed matter physics, and material science.  Thus, low-energy
nuclear reactions at far below the Coulomb barrier should be
explored more in various conditions, experimentally as well as
theoretically.  Finding the method to describe the screening may
eventually help understanding processes in stellar or terrestrial
plasma.


\end{document}